ARTICLE

# Understanding high photocatalytic activity of the TiO$_2$ high-pressure columbite phase by experiments and first-principles calculations

Jacqueline Hidalgo-Jiménez[a,b], Taner Akbay[c], Tatsumi Ishihara[a,d,e] and Kaveh Edalati[a,b,e,*]

The clean production of hydrogen as a zero-emission fuel can be done using photocatalysis, with TiO$_2$ being one of the most promising photocatalysts. However, the activity of TiO$_2$ anatase and rutile phases is still limited. In this study, an oxygen-deficient high-pressure phase of TiO$_2$, columbite, is stabilized by a high-pressure torsion method. The phase is utilized as an active photocatalyst for hydrogen production, and the mechanism of its high activity is examined using density functional theory (DFT). The activity of columbite appears to be experimentally higher than that of the anatase phase. DFT calculations revealed that columbite does not have a narrow electronic bandgap, but its optical bandgap and light absorbance are improved by oxygen vacancies more significantly compared to anatase. Moreover, the water adsorption energy is higher and the surface activation energy for water splitting on the (101) atomic plane of columbite is lower than that for the active planes of anatase. In conclusion, although columbite is not a low-bandgap semiconductor, its large light absorbance and high surface catalytic activity make it a promising candidate for photocatalytic reactions.

## Introduction

CO$_2$ emission due to rapid economic growth has become a major problem in the last few decades. Some of the most polluting activities in our daily life are related to the use of fossil fuels such as in transportation and energy conversion fields. In order to reduce CO$_2$ emissions, hydrogen can play a crucial role by reducing the use of fossil fuels[1]. However, hydrogen is mainly produced by steam reforming which generates a significant amount of CO$_2$ during the process. Therefore, photocatalysis has been considered one of the most promising alternatives for hydrogen production[2].

Among many semiconductor photocatalysts, TiO$_2$ is the first and one of the most efficient[3]. In TiO$_2$ photocatalysis, the oxidation reaction of water for hydrogen evolution occurs by photogenerated charge carriers[3]. Rutile and anatase, with tetragonal crystal structures (Fig. 1(a))[4], are the main TiO$_2$ phases utilized for photocatalysis[1-3]. However, the wide bandgap of these phases being ~3.0 eV for rutile and ~3.2 eV for anatase (i.e. difficult electron-hole generation) and their easy electron-hole recombination limit their photocatalytic performance[5]. TiO$_2$ has a high-pressure columbite phase, with an orthorhombic crystal structure (Fig. 1(a)) which has received

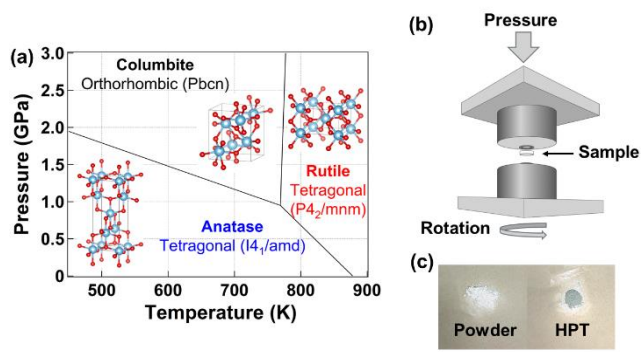

Fig. 1. (a) Phase diagram of TiO$_2$, (b) schematic representation of HPT and (c) appearance of TiO$_2$ pristine powder and sample processed by HPT.

limited attention as a catalyst due to the difficulty in stabilizing it at ambient pressure[6,7].

In this study, a large fraction of the columbite phase is experimentally stabilized by a high-pressure torsion (HPT) process (Fig. 1(b))[8]. The high photocatalytic activity of the high-pressure phase for hydrogen production is experimentally shown and the mechanism of high photocatalytic activity is theoretically examined by using density functional theory (DFT) calculations for the first time.

## Materials and Methods

Both experiments and DFT calculations were employed in this study to clarify the impact of the columbite high-pressure phase on photocatalytic activity.

[a.] WPI, International Institute for Carbon-Neutral Energy Research (WPI-I2CNER), Kyushu University, Fukuoka, Japan
[b.] Graduate School of Integrated Frontier Sciences, Department of Automotive Science, Kyushu University, Fukuoka, Japan
[c.] Department of Materials Science and Nanotechnology Engineering, Yeditepe University, Istanbul, Turkey
[d.] Department of Applied Chemistry, Faculty of Engineering, Kyushu University, Fukuoka, Japan
[e] Mitsui Chemicals, Inc. - Carbon Neutral Research Center (MCI-CNRC), Kyushu University, Fukuoka, Japan
* Corresponding author: Kaveh Edalati (E-mail: kaveh.edalati@kyudai.jp)

## Experimental Procedures

### Sample preparation

It was attempted to generate columbite as much as possible using the HPT method. The initial powder utilized was $TiO_2$ anatase powder (99.8%). About 260 mg of powder was first pressed under 30 kN to generate a 10 mm disc which was later processed by HPT under a pressure of 6 GPa at room temperature. The rotation applied for HPT was 15 turns and the rotation speed was 1 rpm. The resulting material, which had a disc shape with a 10 mm diameter and 0.8 mm thickness, was then crushed using a mortar and pestle, and used for characterization and photocatalytic tests. The appearance of the sample before and after HPT processing is shown in Fig 1(c), indicating that the color of the HPT-processed sample is gray which is an indication of the formation of color centers such as oxygen vacancies[9].

### Characterization

The phase transformation and crystal structure changes were examined by X-ray diffraction (XRD) using Cu Kα radiation, followed by Raman spectroscopy using a laser source of 532 nm wavelength.

The microstructure of the $TiO_2$ samples was studied by scanning electron microscopy (SEM) using an acceleration voltage of 15 keV and transmission electron microscopy (TEM) using an acceleration voltage of 200 keV. Bright-field (BF), selected area electron diffraction (SAED) and high-resolution imaging techniques were utilized to study the microstructural features.

The formation of oxygen vacancies was evaluated by X-ray photoelectron spectroscopy (XPS) using a radiation source of Mg Kα.

The optical characteristics were examined by photoluminescence (PL) emission spectroscopy with a 325 nm laser source. UV-vis spectroscopy was utilized to study the light absorbance of the samples, followed by a bandgap estimation using the Kubelka-Munk theory[10]. XPS was used to estimate the top of the valence band, and finally, the band structure was studied by combining the UV-vis spectroscopy and XPS data, as described in detail in an earlier publication[9].

### Photocatalytic Activity Measurement

The photocatalytic activity for hydrogen production from water decomposition on the $TiO_2$ powder and HPT-treated sample was analyzed under a full light arc of a 300 W UV Xe source. For the sample preparation, 50 mg of the sample, 27 mL of $H_2O$, 3 mL of methanol (as a sacrificial agent), and 0.25 mL of $H_3PtCl_6 \cdot 6H_2O$ (0.01 M) were mixed. The photocatalytic test was conducted for a single cycle of 180 minutes, in which the amount of hydrogen was measured every 30 minutes.

### Computational Methods

To clarify the possible mechanisms underlying the photocatalytic activity of the columbite phase, three main points were theoretically investigated: (1) the electronic bandgap of the columbite phase of $TiO_2$, (2) the effect of oxygen vacancies in columbite on the reduction of the optical bandgap, and (3) the surface activity of the columbite phase of $TiO_2$ for water splitting. These points were studied by first-principles calculations using density functional theory (DFT). All calculations were performed in the plane wave DFT formalism as implemented in the Vienna *Ab-initio* Simulation Package (VASP)[11,12]. Generalized gradient approximation (GGA) was utilized together with the Perdew-Burke-Ernzerhof (PBE) functional to describe the exchange correlation. Projected-augmented wave (PAW)[13] pseudo potential described 12 valence electrons for titanium ($3p^6$ $4s^2$ $3d^4$), 6 electrons for oxygen ($2s^2$ $2p^4$) and 1 electron for hydrogen. The wave functions were expanded with a cutoff energy of 520 eV. The Brillouin zone was sampled with an optimized number of K points according to the geometric nature of the structure. The graphical visualization was made by utilizing VESTA software[14].

### Estimation of Hubbard U-Term

Since DFT underestimates the bandgap due to coulomb interaction effects, the Hubbard *U* parameter was included in the calculations. For the selection of the *U* value, five $TiO_2$ polymorphs, rutile (tetragonal, P42/mnm), anatase (tetragonal, I41/amd), brookite (orthorhombic, Pcab), columbite (orthorhombic, Pbcn) and baddeleyite (monoclinic, P21/c), were evaluated with different *U* values. For these calculations, all structures were made of 12 atoms (4 Ti and 8 O). Then, the obtained bandgaps were compared with the reported experimental results for rutile, anatase and brookite. These comparisons suggested that *U* = 8 eV is a reasonable value for all polymorphs.

### Oxygen Vacancy Effect on Optical Bandgap

Supercells of anatase and columbite with 96 atoms (32 Ti and 64 O) were studied with different concentrations of oxygen vacancies: 3% (2 out of 64 O atoms were removed) and 6% (4 out of 64 O atoms were removed). These supercells were optimized before the self-consistent field (SCF) calculations and used for the calculation of density of states (DOS), band structure and vacancy formation energy. The k-point grid utilized for anatase and columbite supercells were 8×8×6 and 8×8×8, respectively.

### Surface Activation Energy for Water Splitting

Several columbite surface slabs (atomic planes) were generated by utilizing the code Surfaxe[15] with 15 Å of vacuum and 10 Å of bulk and with zero net electrostatic dipole moments. From several slabs, (011) and (101) atomic planes were selected because of their reported high activity in other $TiO_2$ polymorphs[16–19]. The (011) surface termination consisted of four-coordinated and five-coordinated titanium atoms and two-coordinated and one-coordinated oxygen atoms. Comparatively, the (101) surface termination consisted of five-coordinated titanium and one-coordinated and three-coordinated oxygen atoms. These optimizations were performed utilizing the electronic convergence criteria of $1.0×10^{-5}$ eV and the ionic relaxation force criteria of 0.03 eV/Å. Both Brillouin zones were sampled with a 4×4×1 k-point grid.

On the (011) and (101) surface slabs, the water molecule was set above a five-coordinated titanium atom placed near the center of the slab to perform the first optimization calculation (molecular adsorption). Then, for the dissociation mechanism, the closest one-coordinated oxygen was selected to bond with the nearest hydrogen that was separated from the water molecule, and optimization calculations were carried out. Finally, the nudged elastic band (NEB) and climbing image nudged elastic band (CiNEB) methods[20] were used for finding saddle points and the minimum energy path between the molecular adsorption and dissociative states. The number of intermediate steps along each reaction pathway was set to three. Density functional perturbation theory (DFPT) was also employed to study the phonon dispersion curves and the dynamic stability of the saddle points structures by using the code Phonopy[21].

## Experimental Results

### Phase Transformations

HPT can generate multiple changes in the microstructure of processed materials[8,22]. One of them is the stabilization of high-pressure phases, such as columbite[4]. XRD patterns of the powder and the HPT-processed sample are shown in Fig. 2(a). The initial powder is composed mainly of the anatase phase with a small amount of the rutile phase (~3 wt%). The XRD pattern for the HPT-processed sample shows a significant increase in the columbite fraction. Phase analysis using the PDXL2 software indicates that the fraction of the anatase, rutile and columbite phases after HPT processing reaches ~2 wt%, ~29 wt% and ~69 wt%, respectively. Peak broadening also occurs after HPT processing which is related to the reduction in the crystallite size and the formation of crystal defects such as dislocations[22]. The crystallite size refinement permits the columbite phase to remain stable after releasing the pressure because of an increase in the energy barrier for the nucleation of rutile or anatase phases[6].

In Fig. 2(b), the differences between the center and the edge of the sample processed by HPT are compared with those of the initial powder using Raman spectroscopy. The initial powder shows only peaks for the anatase phase. For the HPT-treated disc, the anatase phase appears near the center; however, the anatase traces almost disappear when the measurements are taken close to the edge. Near the edge, the peaks related to columbite, and rutile phases became more intense because the shear strain is much higher in this region[8].

### Microstructure Evolution

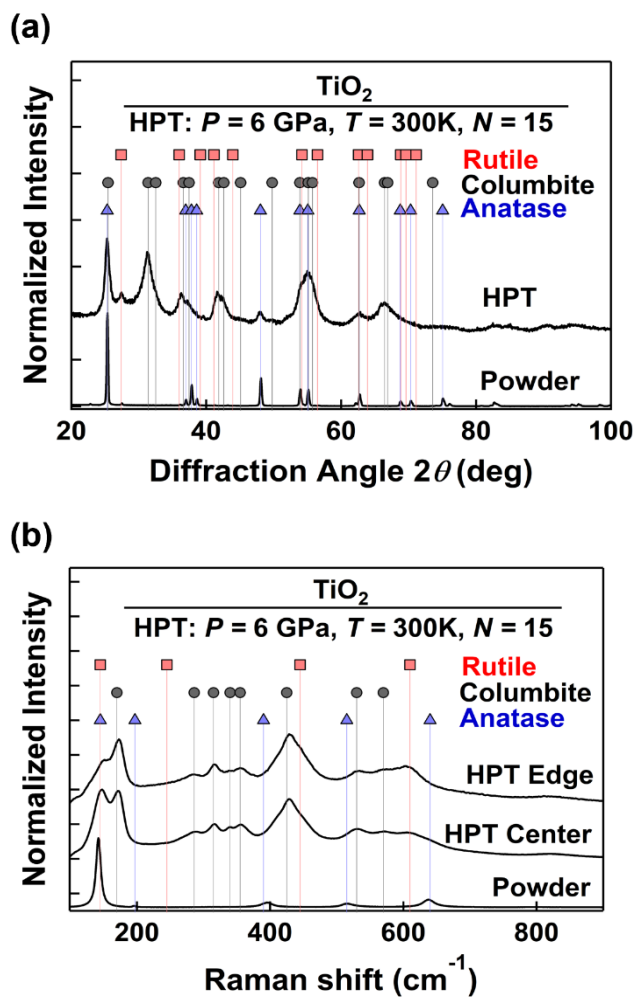

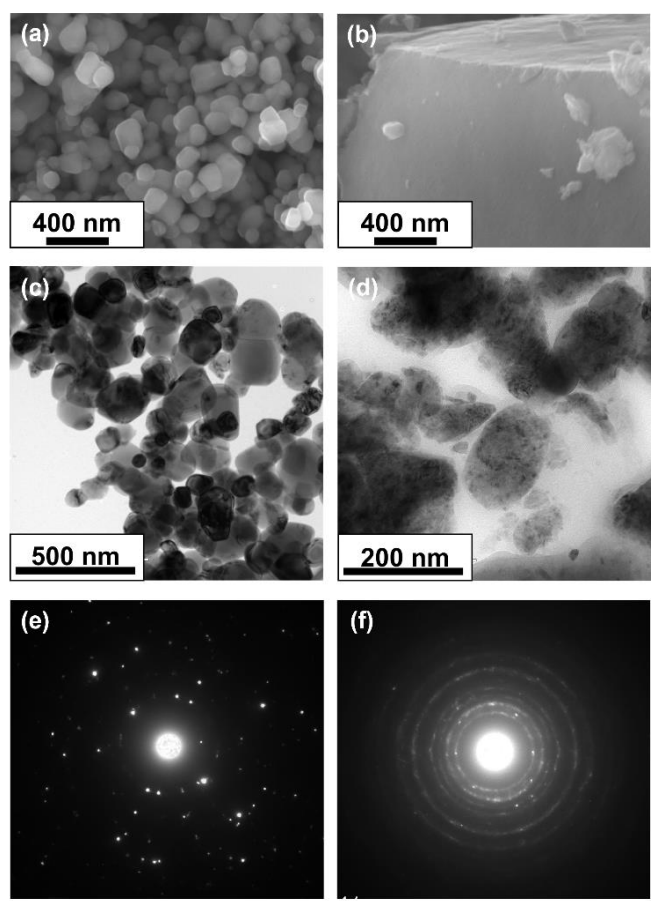

Fig. 2. Columbite phase formation by HPT processing. (a) XRD patterns and (b) Raman spectra obtained for $TiO_2$ after processing in comparison with initial anatase powder.

Fig. 3. Partial consolidation and crystal size reduction in $TiO_2$ after HPT processing. (a, b) SEM micrographs, (c, d) TEM-BF images and (e, f) SAED patters for (a, c, e) $TiO_2$ powder and (b, d, f) sample processed by HPT.

SEM micrographs are shown in Fig. 3 for (a) anatase powder and (b) the sample processed by HPT. These micrographs show the different characteristics of both samples and confirm the partial consolidation of the powders as a result of the HPT process[23]. Based on the SEM images, the average particle size of samples was determined to be 114 nm for the anatase powder and 14.0 µm for the HPT-processed sample. Consolidation of oxide powders by HPT is a consequence of high applied pressure and concurrent shear strain, as discussed earlier for various oxides[23].

Fig. 3 shows the microstructural characteristics of (c, e) initial powder and (d, f) the HPT-processed sample examined in (c, d) TEM-BF and (e, f) SAED modes. Despite the partial powder consolidation by HPT, some dark contrasts appear within the particles which should be due to the formation of nanocrystals and/or lattice defects. The change in the shape of the SAED analysis pattern from a dotted pattern to a ring pattern also demonstrates the reduction in the crystal size from the original powder to the HPT-processed sample. The formation of nanograins is due to the effect of severe plastic deformation induced by the HPT process[8,22].

The HR-TEM images in Fig. 4 were used to examine the nanostructural features of the samples more clearly. These images confirm the presence of the high-pressure phase columbite in the HPT-processed sample, as shown in Fig. 4(a). Additionally, a large amount of grain and interphase boundaries, such as the one in Fig. 4(b), are formed due to the reduction in grain size to the nanometer level. Furthermore, the examination of the micrographs confirms the formation of dislocation-based defects such as Lomer-Cottrell locks, as shown in Fig. 4(c). This type of defect occurs particularly in crystalline materials when a dislocation encounters another dislocation[24]. The presence of these nanograins and dislocations is in good agreement with the XRD peak broadening observed in Fig. 2(a). The formation of boundaries between the $TiO_2$ phases is expected to act as heterojunctions for easy electron-hole separation during light irradiation[25,26].

XPS measurements were used to examine the formation of oxygen vacancies, as shown in Fig. 5 for (a) Ti $2p_{1/2}$ and (b) O 1s. After HPT processing, the spectra for Ti $2p_{1/2}$ show a small extension to lower binding energies while the spectra for O 1s slightly extend to higher energies. These changes confirm that oxygen vacancies are formed after HPT processing. The formation of oxygen vacancies is consistent with the gray color of the HPT-processed sample in Fig. 1c. It should be noted that the concentration of vacancies can be enhanced in the HPT process by either increasing the applied strain (i.e. increasing the number of rotations)[27] or increasing the processing temperature[9,28].

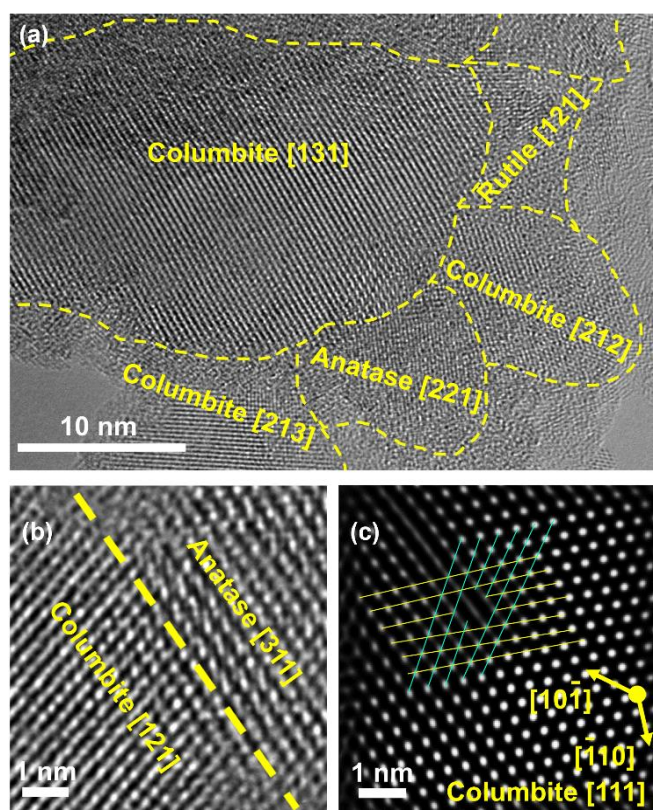

Fig. 4. Formation of nanocrystals and crystal lattice defects in $TiO_2$ after HPT processing. (a) TEM-HR micrograph, (b) lattice image of the phase boundary, and (c) lattice image of dislocation Lomer-Cottrell lock defect reconstructed using the inverse fast Fourier transform for the HPT-processed sample.

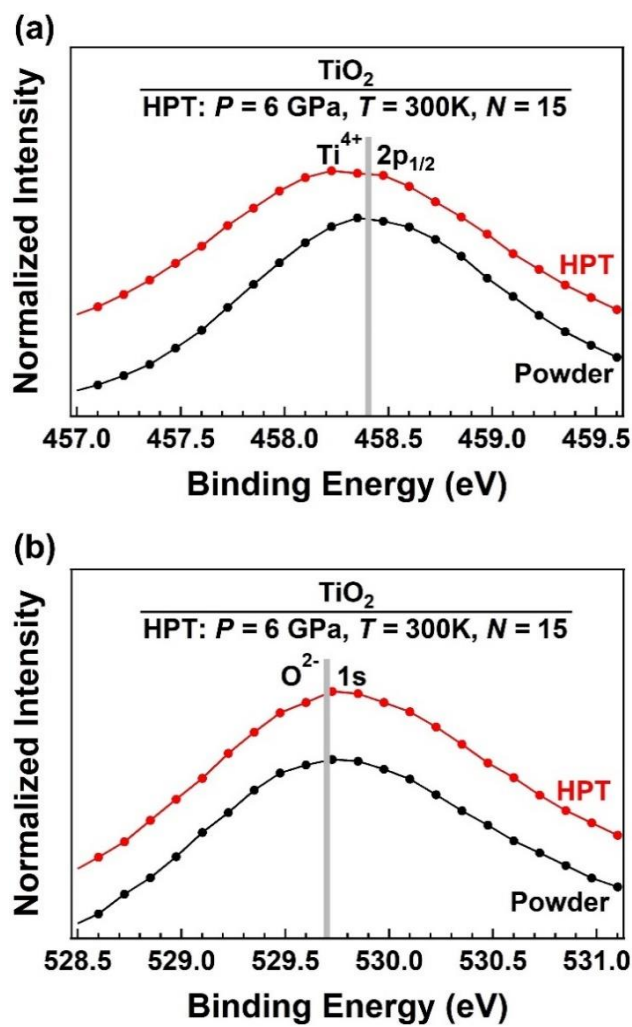

Fig. 5. Formation of oxygen vacancies in $TiO_2$ after HPT processing. XPS profiles of (a) Ti $2p_{1/2}$ and (b) O 1s after HPT processing in comparison with initial anatase powder.

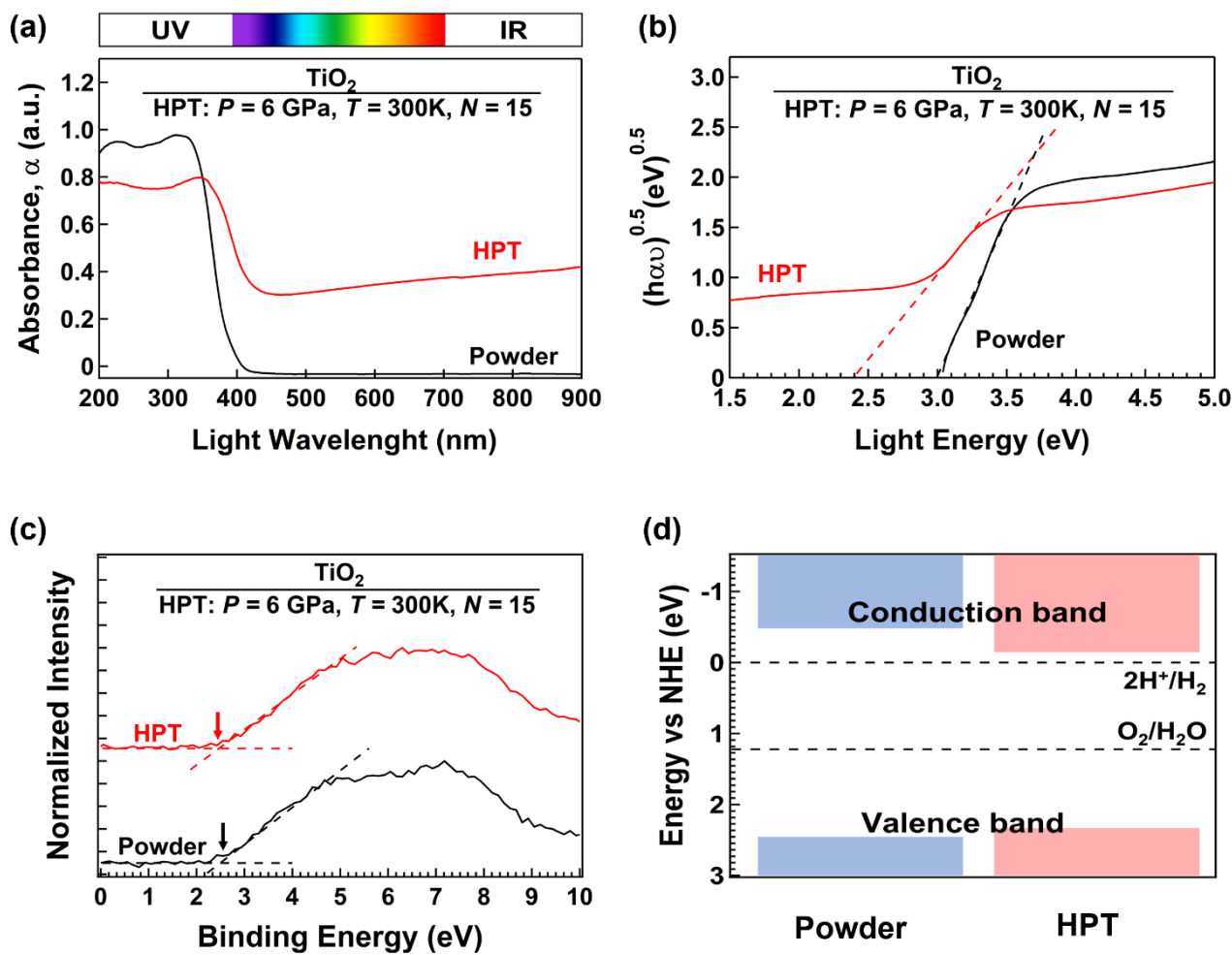

Fig. 6. Enhancement of light absorbance and optical bandgap reduction after HPT processing. (a) UV-vis spectroscopy spectra, (b) Kubelka-Munk analysis for bandgap calculation, (c) XPS spectra for valence band top calculation and (d) band structure obtained for HPT-treated $TiO_2$ in comparison with initial anatase powder.

**Optical Properties**

To examine the optical properties of samples, multiple methods were utilized. First, UV-vis spectroscopy and the Kubelka-Munk method were employed to analyze the light absorbance and the bandgap of the material. The UV-vis spectra are shown in Fig. 6(a), indicating that the light absorbance is higher for the HPT-processed sample. Moreover, as shown in Fig. 6(b), $TiO_2$ exhibits a reduction in the optical bandgap after HPT processing. The initial powder has a bandgap of 3 eV, while the bandgap of the sample processed by HPT is 2.4 eV. The enhancement in the light absorbance after HPT processing was also reported in other HPT-processed oxides and attributed to phase transformations and/or lattice defect generation[9,27,28].

Both samples were analyzed by XPS in order to understand their valence band structure. Fig. 6(c) shows the XPS data to determine the top of the valence band. The valence band top for the powder and the HPT-processed sample is 2.5 and 2.3 eV vs NHE, respectively. By considering the bandgap and the valence band top energy, the bottom of the conduction band is calculated to be -0.5 eV vs NHE for the anatase powder and -0.15 eV vs NHE for the HPT-processed sample. The electronic band structure is summarized in Fig. 6(d), showing that both powder and HPT-processed samples satisfy the thermodynamic requirements for water splitting[1-3].

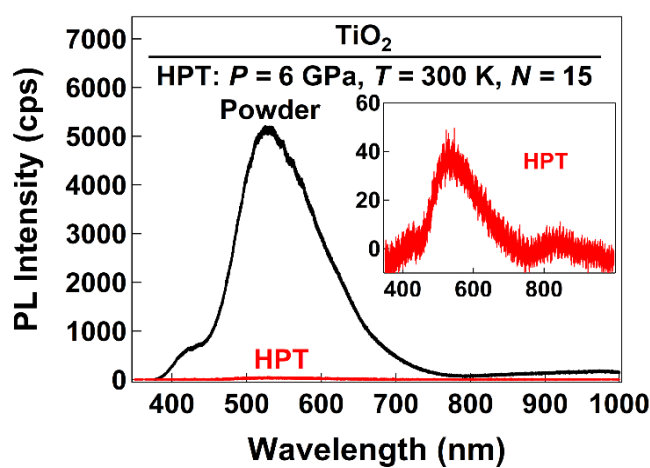

Fig. 7. Reduction of charge carrier recombination after HPT processing. PL emission spectra for $TiO_2$ after HPT processing in comparison with initial anatase powder.

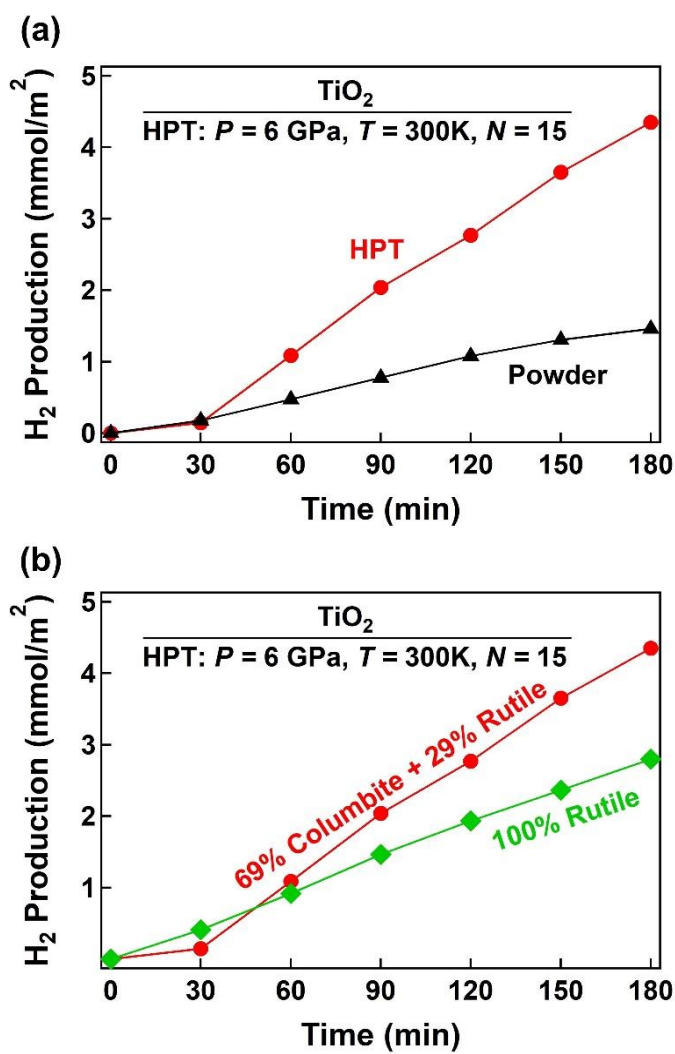

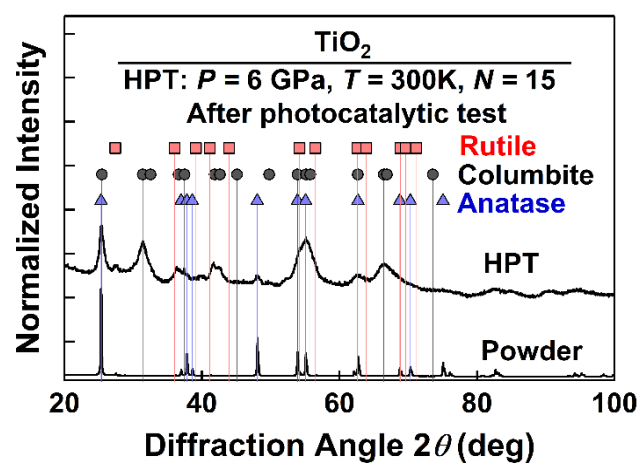

Fig. 9. Stability of TiO$_2$ photocatalysts. XRD profiles after the photocatalytic test for anatase powder and HPT-processed TiO$_2$.

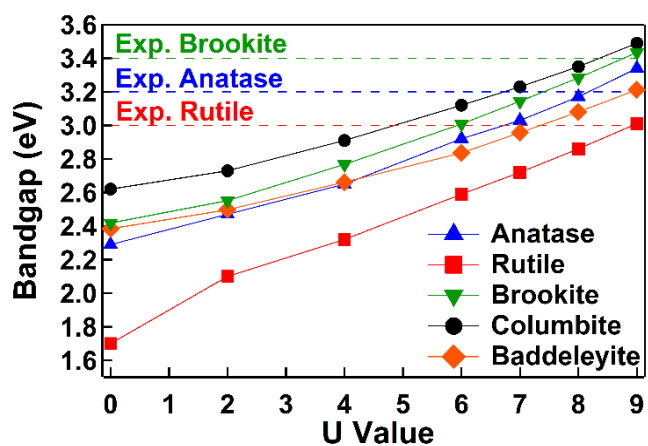

Fig. 8. Enhancement of photocatalytic hydrogen evolution after HPT processing. Hydrogen produced per surface area of the catalyst versus irradiation time under UV light for (a) anatase powder before and after HPT processing and (b) HPT-processed anatase (containing 69 wt% columbite, 29 wt% rutile and 2 wt% anatase) and rutile (containing 100% rutile).

Fig. 10. Large electronic bandgap of the high-pressure columbite phase. Variation of the electronic bandgap with $U$ values of 0 to 9 for anatase, rutile, brookite, columbite and baddeleyite. Horizontal lines at 3.0, 3.2 and 3.4 eV indicate reported experimental bandgaps for rutile, anatase, and brookite, respectively.

The PL spectroscopy measurement was utilized to study the recombination of the charge carriers. As shown in Fig. 7, the PL intensity drastically decreases after the HPT process. For the initial powder, one peak at ~510 nm is detected which is associated with oxygen vacancies and shallow bulk traps of anatase[7,29]. These results indicate that the radiative recombination of the carriers is significantly decreased by the utilization of HPT, which is a positive feature for photocatalysis[1-3].

## Photocatalytic Activity

The photocatalytic activity of samples was tested for hydrogen production under the full arc of Xe light. Fig. 8(a) shows a comparison of the photocatalytic activity of the initial powder and the HPT-processed sample. The photocatalytic activity was normalized by surface area, calculated from the particle sizes obtained by the SEM analysis, in order to have a reasonable comparison. The HPT-processed sample has higher activity than the initial powder, and the photocatalytic hydrogen production obtained for this sample is almost 4 times better than that of the initial powder. These results show that the stabilization of columbite with lattice defects such as oxygen vacancies and dislocations can improve the photocatalytic activity. Since the HPT-processed sample contains not only the columbite phase but also 29 wt% of the defective rutile phase, an additional photocatalytic test was conducted on rutile processed by HPT for 15 turns under 6 GPa and at room temperature to confirm whether the high activity is due to the presence of defective columbite or rutile. As shown in Fig. 8(b), the defective rutile phase, which did not show any phase transformation by HPT, has good activity, but its activity does not reach the activity of the sample containing 69 wt% of columbite. It is concluded that the formation of the defective columbite phase is the main reason for the improvement of photocatalytic activity in this study. Such improvement should be due to higher light absorbance, a lower electron-hole recombination rate and higher fractions of active sites (e.g. heterojunctions, dislocations or vacancies)[6,7].

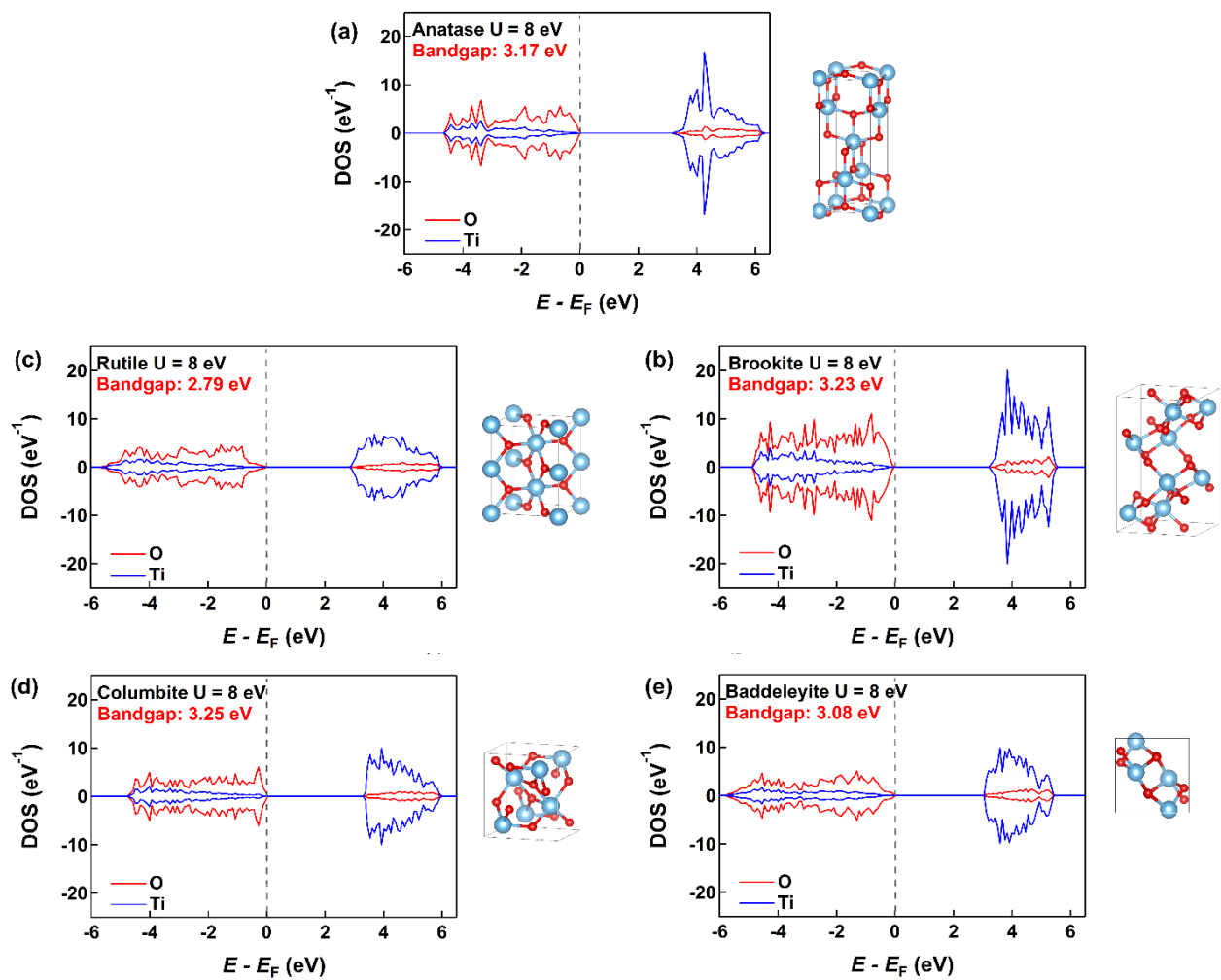

Fig. 11. Contribution of oxygen to the valence band top and of titanium to the conduction band bottom in TiO$_2$ polymorphs. Calculated DOS using a *U* value of 8 eV for (a) anatase (b), rutile (c) brookite, (d) columbite and (e) baddeleyite.

The stability of the material was also evaluated by XRD after the photocatalytic test. The XRD results are plotted in Fig. 9. These XRD profiles show no changes compared to the previously shown patterns in Fig. 2(a) before the photocatalytic test, demonstrating the stability of the high-pressure phase under UV light. The high chemical stability is a positive feature in long-term utilization of photocatalysts for different reactions such as water splitting or CO$_2$ conversion[5].

## Computational Results

To understand the high light absorbance and high photocatalytic activity of the HPT-processed sample, three main points were examined using DFT calculations. First, the bandgap of the columbite phase was compared with that of other TiO$_2$ polymorphs. Second, the effect of oxygen vacancies in columbite on the reduction of the optical bandgap (i.e. enhanced light absorbance) was computed. Third, the surface activity of the columbite phase for water splitting was examined and compared with the activity of anatase and rutile reported in the literature.

### Electronic Bandgap for Different Polymorphs

DFT is an accurate method to carry out a comparative study of the electronic structure of materials; however, DFT tends to erroneously underestimate the bandgap due to delocalized charge distribution[30–32]. To deal with this underestimation, DFT+U was used in this study for the estimation of the bandgap of columbite. The DFT+U results were first compared with the experimentally known bandgaps for anatase (3.2 eV), rutile (3.0 eV) and brookite (3.4 eV)[3,6,33,34] to determine the most reasonable $U$ value. After the selection of the $U$ value, the bandgap of columbite was calculated. A summary of the calculated bandgaps obtained by the variation of $U$ values is shown in Fig. 10, indicating an increase in the bandgap with increasing the $U$ value. As a result of the comparison between the experimental bandgaps and calculated data, the $U$ values that reach the experimental bandgap are 8 eV for anatase and brookite, and 9 eV for rutile. These $U$ values also coincide with the reported values by other researchers for $TiO_2$[32,35]. Therefore, a $U$ value of 8 eV was utilized in all calculations in this study.

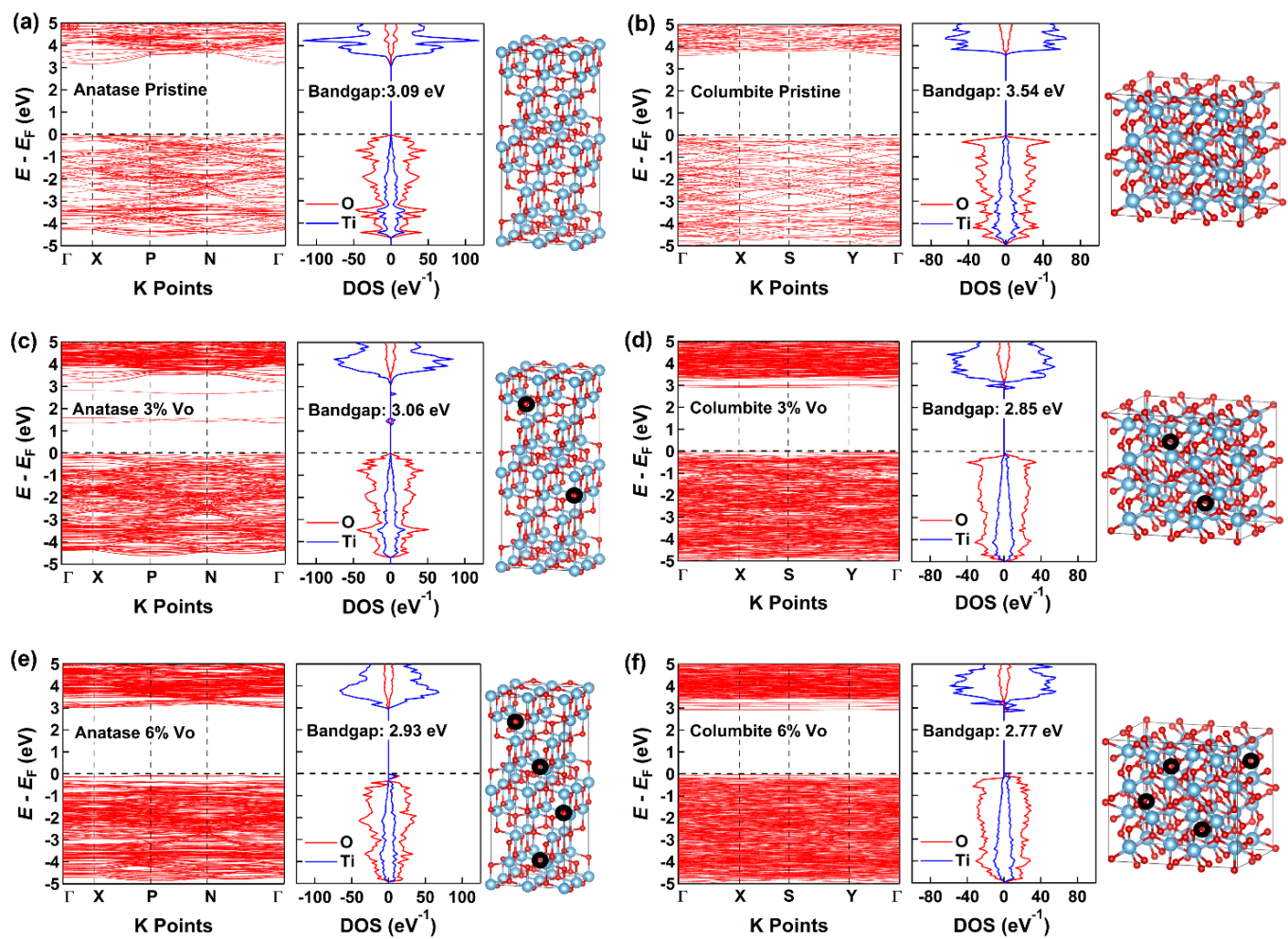

Fig. 12. Reduction of the optical bandgap of columbite by oxygen vacancies. Calculated DOS and the band structure and corresponding modelled supercells using a $U$ value of 8 eV for anatase with (a) 0, (c) 3 and (e) 6 at% of oxygen vacancies and for columbite with (b) 0, (d) 3 and (f) 6 at% of oxygen vacancies. Dotted vertical lines represent the Fermi level ($E_F$) and black circles in supercells show oxygen vacancy positions.

Fig. 11 shows a comparison between DOS plots using a $U$ value of 8 eV. These plots indicate three major points. (i) The valence band is mostly composed of oxygen 2p orbitals while the conduction band is composed of titanium 3d orbitals. (ii) Hybridizing orbitals can be observed near 6 eV. (iii) The bandgap of columbite by utilizing a $U$ value of 8 eV is estimated as 3.4 eV, indicating that columbite is not a low-bandgap semiconductor and its high light absorbance and good photocatalytic activity should be influenced by other factors, as will be discussed below. Moreover, the baddeleyite phase, which was reported as a low-bandgap polymorph in an earlier theoretical study[36], appears to be a wide-bandgap semiconductor as well.

**Oxygen Vacancy Effect on Electronic Structure**

The experimental results showed that the HPT process produces various lattice defects in $TiO_2$ and defects play an important role in different reactions due to the alterations of the electronic structure of the materials. In this section, oxygen vacancies as the most effective defects in photocatalysis are studied using DFT calculations. The effect of different oxygen vacancy concentrations on the DOS and band structure in columbite is compared with that in anatase. For these calculations, 2 and 4 oxygen atoms were randomly removed from the structure representing 3 and 6 at% of oxygen vacancies (Vo), respectively. DOS plots are shown in Fig. 12 for (a, c, e) anatase and (b, d, f) columbite. For anatase with 3% Vo (Fig. 12(c)), two defect states are formed: (i) shallow defect states near the conduction band and (ii) deep defect states. Nonetheless, the deep defect state disappears, and shallow defect states merge with the conduction band by increasing the percentage of oxygen vacancies to 6% (Fig. 12(e)). In columbite with 3 and 6% Vo, only shallow states merged with the bottom of the conduction band are observed. The shallow states merged with the conduction band are the result of $Ti^{3+}$ formation due to extra electrons left by oxygen vacancies[5,16,37,38]. For both anatase and columbite, the optical bandgap decreases by increasing the concentration of oxygen vacancies, but the reduction is more significant for the columbite phase. The presence of oxygen vacancies and $Ti^{3+}$ cations together with narrowing the optical bandgap has a positive effect on the photocatalytic activity of $TiO_2$[16,39].

Table 1. Calculated oxygen vacancy formation energy in bulk and on the surface of anatase and columbite in comparison with reported data in the literature.

| Vo | Phase | Position | Calculation method | $E_{Vacancy}$ (eV) | Ref. |
|---|---|---|---|---|---|
| 12.5% | Anatase | (101) | GGA+$U$ (3.5 eV) | 4.69 | 16 |
| 25% | Anatase | (101) | GGA+$U$ (3.5 eV) | 4.88 | 16 |
| 37.5% | Anatase | (101) | GGA+$U$ (3.5 eV) | 5.04 | 16 |
| 2.08% | Anatase | (101) | GGA+$U$ (6 eV) | 5.26 | 37 |
| 4.16% | Anatase | (101) | GGA+$U$ (6 eV) | 5.14 | 37 |
| 6.25% | Anatase | (101) | GGA+$U$ (6 eV) | 5.28 | 37 |
| 8.33% | Anatase | (101) | GGA+$U$ (6 eV) | 5.06 | 35 |
|  | Anatase | Bulk | GGA+$U$ (4.2 eV) | 4.23 | 39 |
| 5.55% | Anatase | (101) | GGA+$U$ (3 eV) | 4.14 | 40 |
| 12.5% | Anatase | Bulk | HSE06 | 6.6 | 41 |
| 6.25% | Anatase | Bulk | HSE06 | 6.4 | 41 |
| 3.12% | Anatase | Bulk | HSE06 | 6.2 | 41 |
| 1.56% | Anatase | Bulk | HSE06 | 5.2 | 41 |
|  | Anatase | Bulk | HSE | 4.27 | 42 |
|  | Anatase | (101) | HSE | 4.37 | 42 |
|  | Anatase | (001) | HSE | 4.74 | 42 |
|  | Anatase | (101) | HSE06 | 4.81 | 43 |
|  | Anatase | Bulk | HSE06 | 5.46 | 43 |
| 3% | Anatase | Bulk | GGA+$U$ (8 eV) | 5.85 | This study |
| 3% | Columbite | Bulk | GGA+$U$ (8 eV) | 3.15 | This study |

To clarify the thermodynamics of the generation of oxygen vacancies in columbite, its oxygen vacancy formation energy was also calculated by utilizing Equation 1 and compared with that of anatase.

$$E_{Vacancy} = \frac{(E_{Defect} + n\mu_0) - E_{Pristine}}{n} \qquad (1)$$

where $E_{Defect}$ represents the energy calculated using the SCF for the structure with 3% Vo, $E_{Pristine}$ represents the energy for the pristine structure with no defects, $\mu_0$ is the chemical potential and $n$ is the number of vacancies per supercell. The calculated oxygen vacancy formation energy for anatase and columbite is given in Table 1 in comparison with some reported data in the literature[16,37,39-42]. The calculated energy for anatase is in good agreement with the reported data by previous researchers, while columbite has a lower oxygen vacancy formation energy than anatase. When $E_{Vacancy}$ is less positive, the formation of vacancies is energetically more favorable, and thus, columbite should have a higher susceptibility to oxygen vacancy formation.

**Surface Activity for Water splitting**

The calculations for water splitting reactions were performed on the surface of the (101) and (011) atomic planes of columbite. These two surfaces were selected because they were reported as the most active ones for anatase and rutile[16,17,37,44,45]. Water adsorption was performed by attaching a water molecule to a five-coordinated titanium atom and the adsorption energy ($E_{Adsorption}$) was calculated according to Equation 2.

$$E_{Adsorption} = E_{(Slab+H2O)} - (E_{H2O} + E_{Slab}) \qquad (2)$$

where $E_{H2O}$ is the energy calculated for the water molecule in a vacuum, $E_{Slab}$ is the energy of the slab, and $E_{(Slab + H2O)}$ is the total energy of the system after $H_2O$ is absorbed on the slab. For the slab (011), the adsorption energy was +10 eV which implies that the bond between the water molecule and the surface is thermodynamically unfavorable. On the other hand, the adsorption energy calculated for slab (101) was -1.67 eV, suggesting that the columbite slab (101) can absorb the water molecules on the surface. Similar calculations for the anatase slab (101) led to an adsorption energy of -0.96 eV indicating that water adsorption on columbite is more energetically favorable.

Following the water adsorption calculations, reaction pathways for water splitting ($H_2O \rightarrow H+OH$) were calculated by the NEB and CiNEB methods. The NEB and CiNEB calculations for water splitting reactions on slabs (011) and (101) are shown in detail in Fig. 13. In the case of (011) shown in Fig. 13(a), the energy barrier for the reaction is 0.04 eV, and the energy difference between the adsorption and dissociative states is +0.01 eV. Therefore, despite the small energy barrier, the water-splitting reaction on this atomic plane may not be energetically favorable considering that the DFT calculations are conducted at absolute temperature. The energy barrier for slab (101) shown in Fig. 13(b) indicates that the water molecule can follow the proposed reaction path. The energy barrier for the reaction is 0.013 eV and the energy difference from the first to the last step is -0.07 eV, which suggests that the reaction on the slab (101) is thermodynamically exothermic and energetically favorable.

Phonon calculations of the saddle points of the hydrogen evolution reaction on columbite atomic planes (011) and (101) in Fig. 13 are shown in Fig. 14. It is evident that saddle points for both atomic planes show imaginary phonon frequencies, indicating the dynamic instability of the saddle point[21]. These phonon calculations confirm that the catalytic reaction in Fig. 13 goes through a saddle point in which an energy barrier exists between the initial ($H_2O$) and final (H+OH) points of the reaction pathway.

The water-splitting reaction energy barriers calculated for columbite (101) confirm the high surface activity of this high-pressure phase for water splitting. Table 2 presents the values reported for energy barriers for different $TiO_2$ slabs of anatase and rutile[16–18,46–51] and compares them with the columbite phase. Comparatively, the energy barrier for water splitting of columbite slab (101) is smaller than those reported for the active slabs of anatase and rutile. It is concluded that the columbite phase provides high surface activity for the formation of H and OH groups which can accordingly take part in the formation of $H_2$ and $O_2$ molecules in the presence of appropriate co-catalysts.

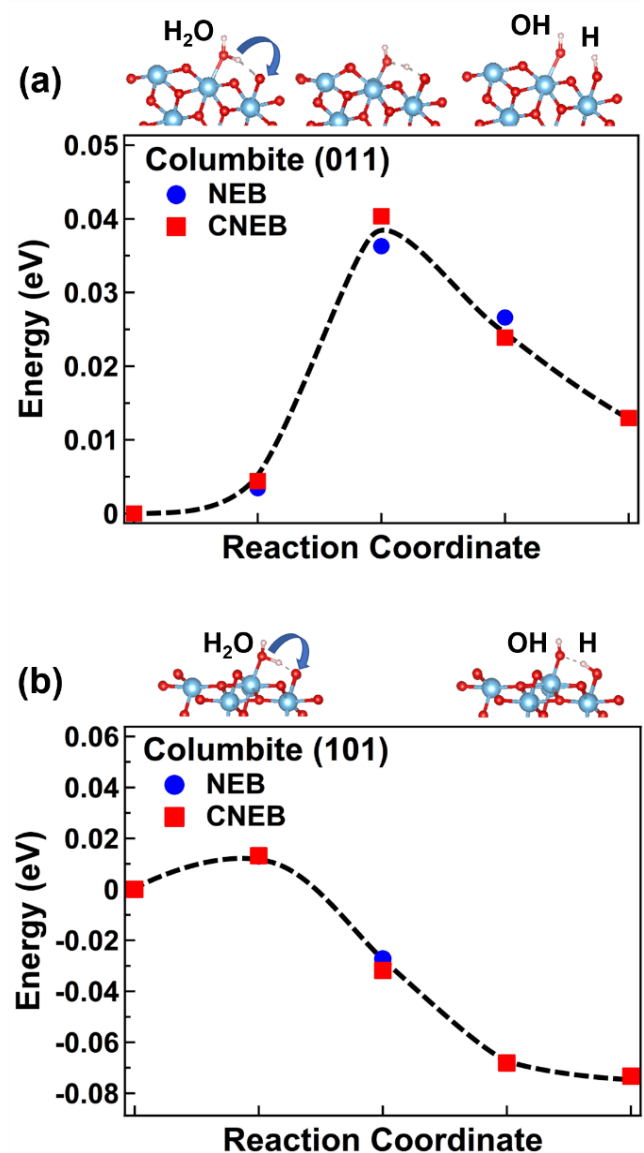

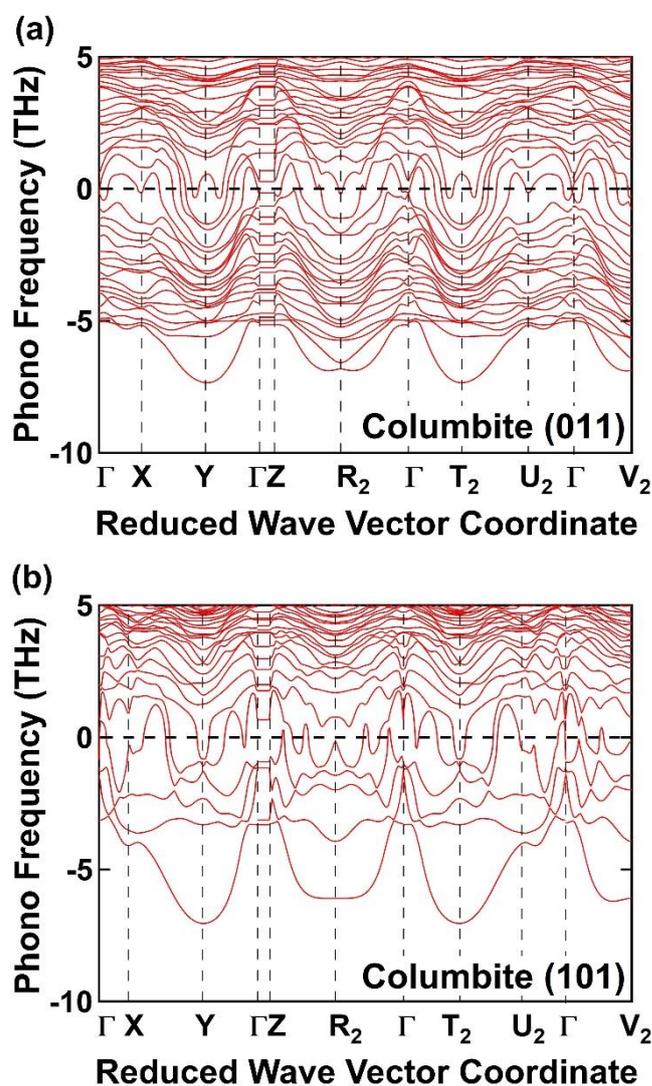

Fig. 13. High surface activity of the (101) atomic plane of columbite for the water splitting reaction. Reaction pathways for water splitting on (a) (011) and (b) (101) atomic planes of columbite calculated with a U value of 8 eV. The pathway for the three main steps of water splitting ($H_2O$→H+OH) on $TiO_2$ is shown on top of each plot.

Fig. 14. Dynamic instability at saddle points of hydrogen evolution reactions in Fig. 13. Phonon dispersion curves at saddle points of the water-to-hydrogen reaction over the columbite atomic planes (a) (011) and (b) (101). Imaginary modes are shown in negative values.

Table 2. Calculated energy difference and energy barrier for the water splitting reaction on the (101) and (011) atomic planes of columbite in comparison with reported data for TiO$_2$ anatase and rutile.

|  | Method | Energy barrier (eV) | Ref. |
| --- | --- | --- | --- |
| Anatase (101) | GGA | 0.02 | 16 |
| Rutile (110) | GGA | 0.16 | 17 |
| Rutile (001) | GGA | 0.109 | 17 |
| Rutile (100) | GGA | 0.225 | 17 |
| Rutile 011 | GGA | <0.1 | 18 |
| Anatase (101) | GGA | 2.67 | 46 |
| Rutile (110) | GGA+$U$ | 0.92 | 48 |
| Rutile (110) | GGA | 0.2 | 49 |
| Rutile (110) | GGA + $U$ | 0.12 | 49 |
| Rutile (110) | GGA | 0.28 | 50 |
| Anatase (101) | GTH* | 0.52 | 51 |
| Anatase (001) | GTH* | 0.05 | 51 |
| Anatase (100) | GTH* | 0.22 | 51 |
| Columbite (011) | GGA | 0.044 | This study |
| Columbite (101) | GGA | 0.013 | This study |

## Discussion

The experiments conducted in this study show that the high-pressure columbite phase is beneficial for the enhancement of light absorbance and the improvement of photocatalytic activity for hydrogen production from water. A large fraction of oxygen-deficient columbite (69 wt%) was stabilized by severe straining via the HPT method, which is larger than those reported in earlier studies[6,7]. Considering these experimental results three possible mechanisms were theoretically investigated: (i) the electronic bandgap of columbite, (ii) the effect of oxygen vacancies on light absorbance and the bandgap, and (iii) the surface activity of columbite for the water splitting reaction.

Based on the findings of the bandgap calculations using different $U$ values, it was possible to determine that the columbite does not have a lower bandgap than other TiO$_2$ polymorphs. These calculations, which were in good agreement with the reported calculations for anatase and rutile[16,36,41,46], do not support an earlier DFT study that suggested high-pressure TiO$_2$ phases as a low-bandgap semiconductor[34]. Therefore, examination of the first mechanism confirms that enhanced light absorbance and good photocatalytic activity of HPT-processed TiO$_2$ cannot be attributed to a decrease in the electronic bandgap. Oxygen vacancies, especially on the surface are considered highly reactive sites which can enhance photon absorption, photogenerated electron-hole separation, chemical adsorption, reactant molecule activation and ionic conductivity[7,16,38–40,46,47,51]. The effect of oxygen vacancies in anatase and rutile was extensively studied earlier[16,37–43,47,52–54], however, few studies were conducted to clarify the effect of oxygen vacancies on the high-pressure columbite phase. The theoretical calculations in this study confirm that oxygen vacancies in both anatase and columbite can narrow the optical bandgap by merging the defect states with the conduction band. However, a higher bandgap narrowing by oxygen vacancies occurs in columbite which has a higher atomic density per volume unit. These computational results (Fig. 12) reasonably justify the experimental results for higher light absorbance and a lower optical bandgap of HPT-processed TiO$_2$ (Fig. 6). It is concluded that a combination of a high-pressure phase and oxygen vacancy generation is essential to achieve good light absorbance in TiO$_2$. Based on the oxygen vacancy formation energy summarized in Table 2, vacancy formation in columbite is energetically more favorable compared to that in the anatase phase[16–18,46-51], and this can be beneficial to develop oxygen-deficient columbite for water splitting.

In addition to light absorbance, surface chemical absorbance and chemical activity are other key factors in improving photocatalytic activity. The computations show that water absorbance at the (101) surface of columbite is energetically favorable. Moreover, the (101) surface of columbite shows a lower energy barrier for water splitting compared to the active surfaces of anatase and rutile, as summarized in Table 1[16–18,46–51]. Taken together, a combination of high light absorbance and chemical absorbance and surface activity is responsible for the high activity of oxygen-deficient columbite. Although oxygen vacancies in bulk may act as recombination centers for charge carriers[6,7,48,55], the presence of surface oxygen vacancies, multiple phases and formation of heterojunctions after HPT processing are expected to contribute to diminishing the recombination of charge carriers[25,26]. Among these features, oxygen vacancy engineering was employed more widely in developing various active photocatalysts[56,57]. Lower electron-hole recombination after HPT processing, which is expected from the low intensity of PL spectra in Fig. 7 and high photocurrent intensity in an earlier work[7], can also contribute to the high photocatalytic activity. Despite the high potential of the oxygen-deficient columbite phase for photocatalytic hydrogen production, its synthesis method needs to be improved in the future to produce a large quantity of the catalyst with a large surface area. Moreover, other strategies used to generate active sites and enhance the photocatalytic activity of different TiO$_2$ polymorphs (e.g. adding dopants, making porosity, generating facets, etc.)[58–60] can be employed in the future to further improve the photocatalytic activity of columbite.

## Conclusions

A combination of quantum mechanical calculations and experiments was used to understand the high photocatalytic activity of the oxygen-deficient high-pressure columbite phase. The main conclusions can be summarized as follows.
(1) The results of bandgap calculations for different TiO$_2$ polymorphs indicate that columbite is not a low-bandgap catalyst.
(2) Calculations suggest that columbite has a lower vacancy formation energy, and oxygen vacancies can decrease its optical bandgap more significantly compared to anatase, a fact that was experimentally confirmed by light absorbance and optical bandgap measurements.
(3) The (101) surface of columbite demonstrates a high tendency for water molecule absorbance and a low activation energy for water splitting to H and OH, a fact that was experimentally confirmed by photocatalytic hydrogen production.

## Conflicts of interest

There are no conflicts to declare.

## Acknowledgements

The authors would like to thank Mr. Yuta Itagoe for assistance in the water splitting experiment, Dr. Saeid Akrami for help in analyzing TEM images, and Prof. Makoto Arita for his contribution to the XPS analysis. The author JHJ acknowledges a scholarship from the Q-Energy Innovator Fellowship of Kyushu University. This study is supported partly by Mitsui Chemicals, Inc., Japan, partly through Grants-in-Aid from the Japan Society for the Promotion of Science (JP19H05176, JP21H00150 & JP22K18737), and partly by the Japan Science and Technology Agency (JST), the Establishment of University Fellowships Towards the Creation of Science Technology Innovation (JPMJFS2132).